\documentclass[oldversion]{aa}
\usepackage{graphicx}
\usepackage{natbib}
\setcitestyle{aysep={}} 
\usepackage{hyperref}
\usepackage{longtable}
\usepackage{lscape}
\usepackage{siunitx}
\usepackage{caption}
\usepackage{subcaption}
\usepackage{color,ulem}
\usepackage{nicefrac}
\newcommand{\teff}{$\mathrm{T_{eff}}$}
\newcommand{\mj}{$\mathrm{M_{J}}$}
\newcommand{\Mp}{$\mathrm{M_{P}}$}

\begin{document}

   \title{Constraints on the nearby exoplanet $\epsilon$ Ind Ab from deep near/mid-infrared imaging limits
\thanks{Based on archival observations from the European Southern Observatory, Chile (Programmes 0102.C-0592 and 60.A-9107).}
}

  % \subtitle{Young low-mass binaries}

   \author{Gayathri Viswanath\inst{1} \and
          Markus Janson\inst{1} \and
          Carl-Henrik Dahlqvist\inst{2}
          Dominique Petit dit de la Roche\inst{3} \and
          Matthias Samland\inst{1} \and
          Julien Girard\inst{4} \and
          Prashant Pathak\inst{3} \and
          Markus Kasper\inst{3} \and
          Fabo Feng\inst{5} \and
          Michael Meyer\inst{6} \and
          Anna Boehle\inst{7} \and
          Sascha P. Quanz\inst{7} \and
          Hugh R.A. Jones\inst{8} \and
          Olivier Absil\inst{2}\fnmsep\thanks{F.R.S.-FNRS Research Associate} \and
          Wolfgang Brandner\inst{9} \and
          Anne-Lise Maire\inst{2} \and
          Ralf Siebenmorgen\inst{3} \and
          Michael Sterzik\inst{3} \and
          Eric Pantin\inst{10}
          }

   \institute{Department of Astronomy, Stockholm University, AlbaNova University Center, 10691 Stockholm, Sweden\\
              \email{gayathri.viswanath@astro.su.se}
        \and
            STAR Institute, Universit{\'e} de Li{\`e}ge, All{\'e}e du Six Aout 19c, 4000 Li{\`e}ge, Belgium
        \and
           European Southern Observatory, Karl-Schwartzschild-Strasse 2, 85748 Garching, Germany
         \and
           Space Telescope Science Institute, 3700 San Martin Dr, 21218 Baltimore, MD, USA
        \and
           Department of Terrestrial Magnetism, Carnegie Institution of Washington, 5241 Broad Branch Road, NW, 20015 Washington, DC, USA
        \and
           Department of Astronomy, University of Michigan, 311 West Hall,1085 S. University Avenue, 48109 Ann Arbor, MI, USA
        \and
           ETH Zurich, Institute for Particle Physics and Astrophysics, Wolfgang-Pauli-Strasse 27, 8093 Zurich, Switzerland
        \and
           Centre for Astrophysics Research, University of Hertfordshire, College Lane, AL10 9AB, Hatfield, UK
        \and
           Max Planck Institute for Astronomy, K{\"o}nigstuhl 17, 69117 Heidelberg, Germany
        \and
           AIM, CEA, CNRS, Universit{\'e} Paris-Saclay, Universit{\'e} Paris Diderot, Sorbonne Paris Cit{\'e}, Gif-sur-Yvette, France           
             }

   \date{Received ---; accepted ---}

   \abstract{
The past decade has seen increasing efforts in detecting and characterising exoplanets by high contrast imaging in the near/mid-infrared, which is the optimal wavelength domain for studying old, cold planets. In this work, we present deep AO imaging observations of the nearby Sun-like star $\epsilon$ Ind A with NaCo ($L^{\prime}$) and NEAR (10-12.5 microns) instruments at VLT, in an attempt to directly detect its planetary companion whose presence has been indicated from radial velocity (RV) and astrometric trends. We derive brightness limits from the non-detection of the companion with both instruments, and interpret the corresponding sensitivity in mass based on both cloudy and cloud-free atmospheric and evolutionary models. For an assumed age of 5 Gyr for the system, we get detectable mass limits as low as 4.4 $M_{\rm J}$ in NaCo $L^{\prime}$ and 8.2 $M_{\rm J}$ in NEAR bands at 1.5$\arcsec$ from the central star. If the age assumed is 1 Gyr, we reach even lower mass limits of 1.7 $M_{\rm J}$ in NaCo $L^{\prime}$ and 3.5 $M_{\rm J}$ in NEAR bands, at the same separation. However, based on the dynamical mass estimate (3.25 $M_{\rm J}$) and ephemerides from astrometry and RV, we find that the non-detection of the planet in these observations puts a constraint of 2 Gyr on the lower age limit of the system. NaCo offers the highest sensitivity to the planetary companion in these observations, but the combination with the NEAR wavelength range adds a considerable degree of robustness against uncertainties in the atmospheric models. This underlines the benefits of including a broad set of wavelengths for detection and characterisation of exoplanets in direct imaging studies.
   }


\titlerunning{Near/mid-infrared imaging of $\epsilon$ Ind Ab}
\authorrunning{G. Viswanath et al.}

   \maketitle
%
%________________________________________________________________

\section{Introduction}
\label{s:intro}

High-contrast imaging with extreme adaptive optics (ExAO) facilities at $JHK$-band wavelengths has yielded many sub-stellar companions to date, including some directly imaged planets \citep[e.g.][]{marois2010,lagrange2010,macintosh2015,keppler2018}. Since giant exoplanets are fairly hot ($\sim$10$^3$~K) at young ages ($\sim$10$^7$~yr) after their formation \citep{spiegel2012,marleau2019}, they emit a large fraction of their bolometric flux at near-infrared (NIR) wavelengths. This fact, along with the relatively low thermal background in the $JHK$ range and the fairly high AO correction quality that can be reached there, is the reason most exoplanets searches -- and most detections -- occur at $JHK$ wavelengths. However, most exoplanets imaged thus far are relatively young, with an age of $\sim$1-100 Myr. Older gas giant planets, at $\sim$Gyr ages, will have typical effective temperatures of $\sim$100--300~K.
%However, the vast majority of planets are older, and therefore colder, than the exoplanets that have been imaged thus far. At $\sim$Gyr ages, gas giant planets will have typical effective temperatures of $\sim$100--300~K. 
At such temperatures, the flux at $JHK$ wavelengths becomes so low that it is out of reach for present-day facilities. For a range of atmospheric properties, a significant flux bump remains at 4~$\mu$m \citep[e.g.][]{allard2001,burrows2006,fortney2008}, which is included in the red end of the $L^{\prime}$ filter. Hence, $L^{\prime}$ is the shortest feasible wavelength band where old and cold planets can be studied. At longer wavelengths, the planet-to-star contrast is often even more favourable, although the thermal background also increases rapidly with wavelength longwards of $L^{\prime}$, posing a considerable observational challenge of its own. As a result, a number of efforts have been made to develop high-contrast imaging for detecting and characterizing planets in $L^{\prime}$ from the ground \citep[e.g.][]{kasper2007,janson2010,quanz2010,absil2013} and in space \citep[e.g.][]{janson2015,durkan2016,baron2018}.

An important goal in exoplanet research is to detect planets both in imaging and with radial velocity (RV) and/or astrometry. This allows us to simultaneously determine a range of properties such as the mass, orbit, luminosity and spectral distribution of the planet, providing a much larger information space than is available with either technique in isolation. For brown dwarfs, a combination of imaging and RV \citep[e.g.][]{crepp2012,crepp2014,peretti2019} or imaging and astrometry \citep[e.g.][]{calissendorff2018} or all three \citep[e.g.][]{brandt2019,grandjean2019,maire2020b,currie2020} has been achieved in a number of cases, encouraging increasing efforts in recent years to use similar approaches to detect and characterise exoplanets \citep[e.g.][]{mawet2019}.
In two cases so far, directly imaged planets have also been observed astrometrically. The planet $\beta$~Pic b \citep{lagrange2010} causes an acceleration of its host star between the \textit{Hipparcos} \citep{perryman1997} and \textit{Gaia} \citep{brown2018} astrometric epochs \citep[]{snellen2018, nielson2020,brandt2020}. In combination with direct monitoring of the planet motion, this yields a dynamical mass estimate of 11$\pm$2~\mj. Additionally, \cite{nowak2020} recently combined direct observation of a second planet in the system, $\beta$~Pic c \citep{lagrange2019}, with RV and astrometric data, constraining its inclination and luminosity to estimate its mass at 8.2$\pm$0.8 \mj. $\beta$~Pic is a young \citep[$\sim$24~Myr, see e.g.][]{bell2015} system and very unusual in the sense that it is also very nearby at 19.44~pc \citep{brown2018} -- most similarly young planetary systems are much more distant, and thus harder to detect astrometrically. This is further emphasized by the fact that they can only be imaged at very large separations, requiring excessively large astrometric baselines for dynamical detection. The most promising overlaps between dynamical and imaging characterization of planets is in very nearby systems; however, such systems are generally relatively old and thus their planets can be expected to be relatively cold, further motivating the need for direct imaging developments in the $L^{\prime}$ band and longwards \citep{heinze2010}.

Located only 3.639$\pm$0.003~pc away \citep{brown2018}, $\epsilon$~Ind A is one of the most nearby Sun-like (K5-type, 0.76~$M_{\rm sun}$) stars \citep[e.g.][]{demory2009}. A common proper motion low-mass companion, $\epsilon$~Ind B, was reported at a wide separation ($\sim$1460~au) in 2003 \citep{scholz2003}, which shortly afterwards was discovered to itself be binary, forming the brown dwarf pair $\epsilon$~Ind Ba and Bb \citep{mccaughrean2004}. Age estimates in the literature for $\epsilon$ Ind A from chromospheric activity and rotation have ranged from $\sim$1 Gyr to $\sim$5 Gyr \citep[e.g.][]{lachaume1999,barnes2007,feng2019}. However, the lower age estimate of 1 Gyr by \cite{barnes2007} is based on an older less accurate value of rotation period for the star, compared to the more recent and reliable estimate of $35.732^{+0.006}_{-0.003}$ days by \cite{feng2019} which points to an older age of $\sim$4.5 Gyr using the same method of gyrochronology \citep{delorme2011} as used by \cite{barnes2007}. This indicates that $\epsilon$ Ind A system is expected to be older in age. Additionally, $\epsilon$~Ind A has been observed to exhibit a long-term RV trend, which was stronger than the secular acceleration caused by the rapid motion of the star on the sky \citep{endl2002,zechmeister2013}. The trend was also too strong to be explained by any dynamical impact of $\epsilon$~Ind Ba/Bb, indicating that an additional object -- most likely a giant planet -- must be exerting a small but significant gravitational pull on the star. Hence, dedicated imaging surveys were performed \citep{geissler2007,janson2009} in order to try to detect the companion, but yielded no detections. This further underlined that the companion must be a planet, since the detection limits excluded brown dwarfs and more massive objects as potential companions. Meanwhile, RV monitoring continued for the object, and was showing increasingly clear curvature, underlining the reliability of the RV solution and greatly enhancing the predictability of orbit and mass for the planet. On this basis, deeper imaging campaigns were planned and executed that are the topic of this article. Recently, a sufficient coverage in both RV and astrometry was reached such that the planet $\epsilon$~Ind Ab could be confirmed, and such that all of its orbital elements could be constrained \citep{feng2019}. The current best-fit mass and semi-major axis of $\epsilon$~Ind Ab are 3.25$^{+0.39}_{-0.65}$~\mj~(at an inclination of ${64.25^{\circ}}^{+13.8}_{-6.09}$) and 11.55$^{+0.98}_{-0.86}$~au.

Here, we present deep AO imaging observations with the 3.8 $\mu$m NaCo $L^{\prime}$ and 10--12.5~$\mu$m NEAR wavelength filters. NaCo \citep{lenzen2003,rousset2003}, which was a workhorse near-infrared instrument at the Very Large Telescope (VLT) for nearly two decades, was recently decommissioned and a new instrument with enhanced 1--5 $\mu m$ imaging capabilities, ERIS \citep[]{eris2018,eris2020}, will see its first light at VLT in $\sim$2021. NEAR \citep{kaufl2018, pathak2021} was an upgrade of VISIR \citep{lagage2004} with AO assistance and a coronagraph added to the mid-infrared camera. NEAR was developed primarily for the purpose of searching for planets in the $\alpha$~Cen system \citep{kasper2019}, but was briefly offered for broader scientific applications in a science verification program executed during the second half of 2019 before also being decommissioned. Between them, the two instruments offer unprecedented contrast and sensitivity in a wavelength range where cool/temperate planets radiate a significant portion of their thermal flux. In the following, we will describe the observational setups in Sect. \ref{s:obs} and the data reduction schemes in Sect. \ref{s:data}. The results will be discussed in Sect. \ref{s:results}, and the conclusions will be summarized in Sect. \ref{s:summary}.

\section{Observations}
\label{s:obs}

\subsection{NaCo observations}
\label{s:obsnaco}

Observations with NaCo of $\epsilon$ Ind A were acquired in October and November of 2018, using the $L^{\prime}$ filter in pupil tracking mode. Originally allocated in service mode, the observations were transferred to designated visitor mode when the visible wavefront sensor (WFS) broke down and a non-standard implementation of the infrared WFS was required due to the high brightness of $\epsilon$ Ind A.  The first $\sim$quarter-night run was executed on Oct 5, but conditions were poor and the data were not usable. This paper will focus on the subsequent four $\sim$quarter-night runs performed on Oct 12 (MJD = 58403), Oct 26 (MJD = 58417), Nov 3 (MJD = 58425) and Nov 4 (MJD = 58426), all of which had a seeing of $\sim1\arcsec$ or less with clear sky conditions.

During the science sequence, 1 minute exposures were performed through single readout blocks of 300 frames (NDIT) already coadded at the detector level, each with an integration time (DIT) of 0.2 seconds. Full-frame readouts were used so that the star could be placed in the center of a `good' quadrant of the detector, avoiding the lower left quadrant which was affected by a large number of unusable pixel stripes. At the time of the observations, the ephemerides of $\epsilon$ Ind Ab were only loosely constrained, theoretically allowing for separations up to $\sim$5$^{\prime \prime}$. Placing the star near the center of the upper left quadrant of the $\sim$27-by-27$^{\prime \prime}$ detector left enough space both relative to the bad quadrant and the detector edges to accommodate detection at such large separations. The expected separation of $\epsilon$ Ind Ab during the NaCo observation epoch (i.e. as on October 26, 2018) was later determined to be 1.37$\arcsec\pm0.16\arcsec$, following the ephemerides calculated by \cite{feng2019}. The field rotation that was obtained for the data taken on Oct 12, Oct 26, Nov 3 and Nov 4 were 29.5$^\circ$, 34.5$^\circ$, 58.5$^\circ$ and 58$^\circ$ respectively. In total across the four usable observing runs, 456 science frames were acquired, of which 8 were visually classified as being of poor quality and discarded. This left 448 co-added frames for the analysis, corresponding to a total useful integration time of 3.73 hours. No dithering was performed during the observations, but individual sky frames were interspersed among the science observations to monitor the thermal background. The observations were obtained in standard (saturated) imaging mode with pupil tracking. The resulting science data were saturated typically out to $\sim6$ pixels (0.16$\arcsec$) from the central star. Non-saturated frames of the primary star were acquired with an integration time of 0.1s and 100 coadds, with the use of the `long' neutral density (ND) filter of NaCo, with a transmission in the $L^{\prime}$-band of 1.8\%\footnote{\url{https://www.eso.org/sci/facilities/paranal/instruments/naco/doc/VLT-MAN-ESO-14200-2761_v94.pdf}}. 

\subsection{NEAR observations}
\label{s:obsnear}

$\epsilon$ Ind A was observed using NEAR in visitor mode in September 2019, utilizing the enhanced sensitivity and PSF (point spread function) contrast obtained via its combination of AO with an AGPM coronagraph \citep{maire2020}. Observations were carried out during three nights on September 14 (MJD = 58740), September 15 (MJD = 58741) and September 17 (MJD = 58743) using the NEAR (10 - 12.5 $\mu$m) filter in pupil-stabilized tracking mode at good weather conditions.
Science exposures were performed with high-frequency DSM (Deformable Secondary Mirror) chopping at a frequency of 8.33 Hz and a chopping amplitude of 4.5$\arcsec$, in combination with nodding. We used an integration time of 60ms, resulting in 2 chop images per nod, that we averaged. The star was positioned behind the coronagraph near the center of the detector with an effective field of view of $\sim20\arcsec\times20\arcsec$ in the Small Field (SF) mode. The expected separation of $\epsilon$ Ind Ab as on September 15, 2019 was 1.02$\arcsec\pm0.19\arcsec$ (3.71 AU) at a position of <ra>=0.903$\arcsec\pm0.204\arcsec$, <dec>$=-0.302\arcsec\pm0.352\arcsec$, relative to the star. 

%Each chop image had an integration time of 60ms and the chop images at each nod position were averaged. 
The science frames for the target were obtained from the difference image between the two nod positions, giving an on-target integration time of 48s for each nod beam position. A total of 246 such nod difference frames were obtained over the three nights, out of which 23 frames were of poor quality due to unfavourable seeing, AO or tracking malfunction and were discarded. A total of 223 science frames were thus used for further analysis, corresponding to a total integration time of 2.23 hours. An image of $\epsilon$ Ind A not obscured by the coronagraph was obtained for analysis from the nod-difference frames by summing the off-center chop beams at the two nod positions and averaging over all such frames. The integration time for this unobscured image of the central star obtained via the above method is the same as the on-target integration time of each nod difference frame in the coronagraphic images. Since the star does not have a very high photon flux rate in the mid-infrared relative to the background, no specific measures were taken during observation to prevent saturation.

\section{Data reduction}
\label{s:data}

\subsection{NaCo reduction}
\label{s:datanaco}

NaCo was installed at the VLT in 2001, so it had been operational for approximately 17 years when the observations were acquired in late 2018, and was de-commissioned less than a year later in 2019. By the time of the observations, NaCo was experiencing several technical difficulties, not least with respect to the detector which had a large number of bad pixels and a rather unstable flat field with variable stripes and similar patterns. This required particular considerations in the data reduction procedure, expanding on previous NaCo $L^{\prime}$ processing \citep[e.g.][]{janson2008}. As we will see, with dedicated processing it was possible to reach a higher image depth than any other observation of the system, including previous NaCo imaging \citep{janson2009}, although not by the same factor that would have been expected if the detector status had been as good as in those previous epochs.

Sky flats\footnote{\url{https://www.eso.org/sci/facilities/paranal/instruments/naco/doc/VLT-MAN-ESO-14200-4038_v0.pdf}} are essentially the only way to get reliable flat field in $L^{\prime}$ with NaCo, and are acquired on a monthly basis at the telescope. Flat field frames acquired during the last years of NaCo operations exhibited variable features such as horizontal stripes occurring at some epochs but not others; quasi-vertical stripes that changed slightly but significantly in morphology between epochs; and diffuse circular or elongated features across varying parts of the detector field. For this reason, we took special care in identifying a well-matching flat to the observational data, visually examining every flat field with relevant settings that had been acquired throughout 2019. A flat field from November (which is close to the epoch of the observations) was found to be the best option, although it had to be filtered from an additive horizontal stripe pattern that occurred in individual frames. This was done row-by-row by calculating the median flux of 50 pixels near the edges of the frame in that row and subtracting the result from each pixel in the row. Each science frame was dark subtracted and flat field corrected. Bad pixels were identified through their deviations from a 5-pixel median box filtered version of the flat field. Unsharp masking was applied to every science frame through a median box filtering with a box size of 20 pixels that was subtracted from the image. This eliminates the low spatial frequencies that dominate the distribution of the thermal background flux, while having a negligible effect on point sources which have a full width at half maximum (FWHM) of $\sim$4 pixels. The exact same procedure was applied to the non-saturated frames, such that any small flux losses from the procedure would be accurately represented when calibrating the contrast in the final images from the unsaturated PSF.

Post-processing of the science frames was carried out using a combination of two different techniques. In the regime close to the star ($<3\arcsec$), we use an automated version of the Regime Switching Model (RSM) algorithm \citep{dahlqvist2021}, which operates based on the temporal evolution of pixel intensity in the de-rotated cubes of residual frames resulting from an optimum combination of different PSF subtraction techniques based on Angular Differential Imaging \citep[ADI; ][]{marois2006}. In a recent data challenge designed to evaluate the relative performance of different high-contrast algorithms, RSM scored very highly both in terms of high true detection rates and in terms of low false detection rates \citep{cantalloube2020}. Iterating over radial distance, the RSM algorithm uses a 2-state Markov chain on the resulting time-series to generate probabilities, for each pixel in an annulus, to be in two regimes; one where the pixel intensity is described purely by residual noise from quasi-static speckle and another where the pixel intensity is described by both residual noise and a model of the planetary signal. 

\begin{figure*}
\centering
\begin{subfigure}{.5\textwidth}
  \centering
  \includegraphics[width=0.96\linewidth]{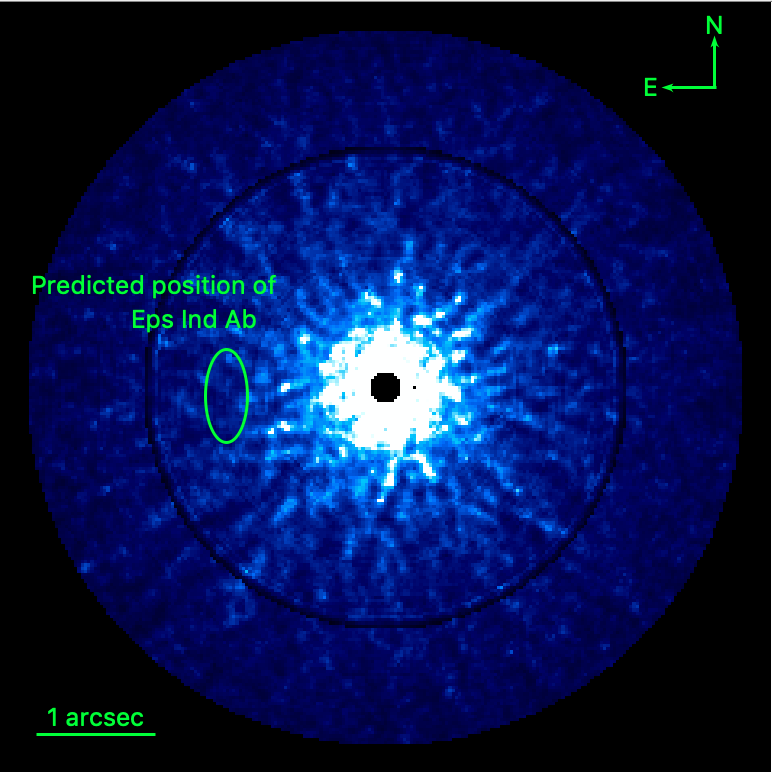} 
  \caption{}
  \label{fig1a}
\end{subfigure}%
\begin{subfigure}{.5\textwidth}
  \centering
  \includegraphics[width=1.\linewidth]{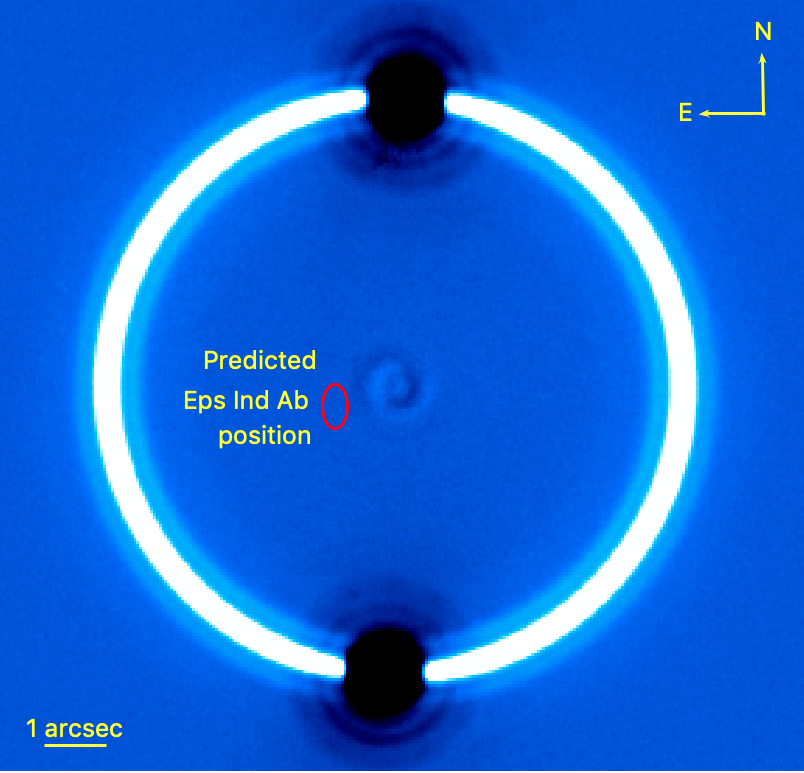}
  \caption{}
  \label{fig1b}
\end{subfigure}
 \caption{(a) The final detection probability map from NaCo and (b) the final reduced image from NEAR observations of $\epsilon$ Ind A. The predicted position of the companion at the time of observation, along with the 1$\sigma$ uncertainty in position, as inferred from the best-fit ephemerides, based on RV and astrometry data \citep{feng2019}, is shown as an ellipse. The NaCo image was reduced using a combination of RSM and modified LOCI subtraction applied at suitable separations while the NEAR image was reduced using the circular profile subtraction technique. No companions were detected in both images.}
\end{figure*}

The RSM algorithm \citep{dalqvist2020a} relies on the off-axis PSF, or a forward-modeled PSF \citep{dahlqvist2021} to account for signal self-subtraction resulting from the reference PSF subtraction, to model the planetary signal. The RSM detection map is then produced by time-averaging the probabilities of being in the planetary regime. In this work we used an enhanced version of the RSM algorithm to generate our final contrasts. This version optimizes the different PSF subtraction techniques used in the algorithm based on minimization of contrast; optimises the algorithm itself based on maximization of probability ratio of injected fake companion to the background noise; and then searches for the best set of PSF-subtraction techniques and observation sequences to generate the optimum contrast. This best set is selected via a bottom-up greedy selection algorithm. The algorithm iteratively adds the PSF-subtraction technique and observation sequence that maximizes the incremental increase of the probability ratio of injected fake companion to the background noise, until no more incremental increase can be obtained.
Consequently, the final detection map shown in Figure \ref{fig1a} has been generated using the local low-rank plus sparse plus Gaussian decomposition \citep[LLSG; ][]{gomez2016} for the first observation sequence, the annular principle component analysis \citep[Annular PCA; ][]{gomez2017} for the second, the locally optimised combination of images \citep[LOCI; ][]{lafreniere2007} for the third and a combination of all three PSF-subtraction techniques for the fourth observation sequence\footnote{The off-axis PSF has been used to model the planetary signal for all three PSF subtraction techniques; the cube of residuals generated with forward-modeled PSF was not selected during the optimization process.} (respectively, $12^{th}$ and $26^{th}$ October, $3^{rd}$ and $4^{th}$ November). No companions were detected in the final map. 

Since the noise statistics in the map are unknown, a standard 5$\sigma$ threshold cannot be defined to generate contrast curves. The contrast curves are instead obtained by injecting fake companions into the observation sequences with different flux values, at a range of azimuths and radial separations. The detection limit is defined as the flux at which the True Positive Rate (TPR) corresponding to the detection of the first false positive in the entire frame ($<3\arcsec$) is 50\% (which is similar to a standard 5$\sigma$ criterion, see \cite{dahlqvist2021} for more details). To add further robustness to our results, we also produce a contrast curve corresponding to a TPR of 90\%. In Figs. \ref{fig2a} and \ref{fig2b}, we show examples of the 50\% and 90\% criteria by injecting 11 fake companions into the observing sequence at the level of the calculated contrast curve, all at a mutual azimuthal angle, along the same radial distances at which the contrasts are estimated. The figures show the detected signals at a threshold equal to the brightest speckle in the empty probability map (0.0025), at a TPR of 50\% and 90\% respectively. Out of the injected companions, 4 are retrieved for the 50\% case and 9 for the 90\% case. The slight deviations from exactly 50\% and 90\% retrievals are due to statistical fluctuations; the numbers quoted here correspond to a single azimuth, while fake companions were injected at 10 different azimuths per angular distance to compute the contrast curves.

As we mentioned above, the RSM algorithm is applied out to a separation of 3$^{\prime \prime}$; in the background-limited regime outside of this range, RSM offers no advantages relative to conventional ADI-based PSF subtraction techniques \citep{dahlqvist2021}. Furthermore, owing to the aforementioned detector noise and instability of the flat field correction, we have found that conventional techniques such as ordinary LOCI perform significantly worse in this regime than expected, given the integration time and predicted background level.

To mitigate this effect, we have performed an alternative LOCI-based subtraction procedure. In this alternative implementation, we ignored the small drift of the stellar PSF that occurred during the observations, and operated on the images as if the star had been perfectly fixed with respect to the detector. In other words, LOCI was applied to the non-shifted individual images, and shifting was applied only after the reduction had finished. This procedure yields a worse performance than the normal procedure in the contrast-limited regime, because the unaccounted relative shifts in the stellar PSF result in a worse average PSF matching. However, the large benefit of the non-shifted implementation is in the (nominally) background-limited regime. When the image are shifted prior to optimization and subtraction, pixels are shifted from their original locations, such that any systematic error that might have occurred at the pixel level gets spread across different parts of the resampled images, forming a quasi-Gaussian noise term in each new pixel with (in this case) a larger variance than the shot noise of the thermal background. By operating on the non-shifted frames, the LOCI subtraction has the chance to subtract out systematic noise effects on the detector level, potentially yielding a better performance outside of the contrast-limited central regime. A 5$\sigma$ contrast curve as function of separation was derived based on the standard deviation of samples within an annulus of a given separation, where each sample was the aperture sum of pixels around the sampling point with a diameter equal to the FWHM, 120 mas. Self-subtraction was accounted for by using the unsaturated stellar image as a PSF representation and calculating the flux loss at every location in the image based on the actual optimization coefficients used in the modified LOCI subtraction. For the final output, we use the modified LOCI implementation from 3$^{\prime \prime}$ and outwards, which is outside of the region used for the optimization between 2.7$^{\prime \prime}$ and 3$^{\prime \prime}$. Hence, there is no systematic subtraction of any hypothetical companions from the optimization in the region of interest, and therefore the only recurring source of self-subtraction comes from partial PSF overlaps from each pair of companion signatures in any pair of mutually subtracted images. Due to the large separation of $>$3$^{\prime \prime}$, such overlaps are small, and as a result, self-subtraction is low in the part of the special LOCI output image used in this analysis.

The non-shifted procedure did indeed perform very well in the outer ranges of the image, with a substantially better contrast curve than the shifted procedure in that regime, as can be seen from Figure \ref{comp_con}, which shows the comparison of contrast curves obtained using different reduction techniques used on NaCo L$^{\prime}$ data in this work.

\begin{figure}
  \centering
  \includegraphics[width=1.05\linewidth]{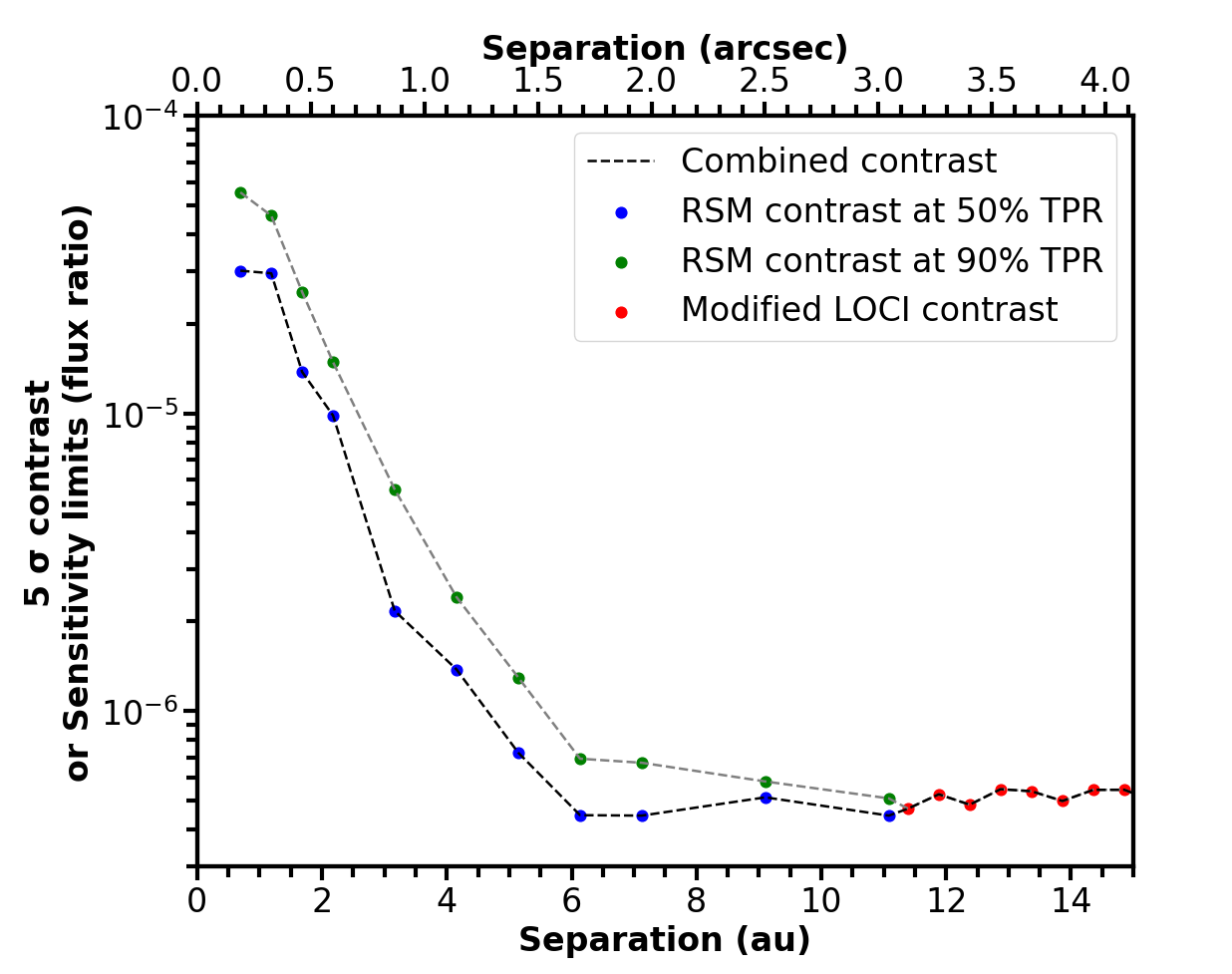}
  \caption{The composite contrast curve from NaCo L$^{\prime}$ observations, obtained from the combination of RSM and the modified LOCI subtraction techniques. The y-axis shows the 5$\sigma$ contrast limits obtained from modified LOCI subtraction techniques (shown in red solid dots), beyond $\sim$11 au and the equivalent flux ratio or sensitivity limits obtained using RSM algorithm corresponding to a TPR of 50\% (blue solid dots) and 90\% (green solid dots). }
  \label{fig3a}
\end{figure}

\begin{figure}
  \centering
  \includegraphics[width=1.05\linewidth]{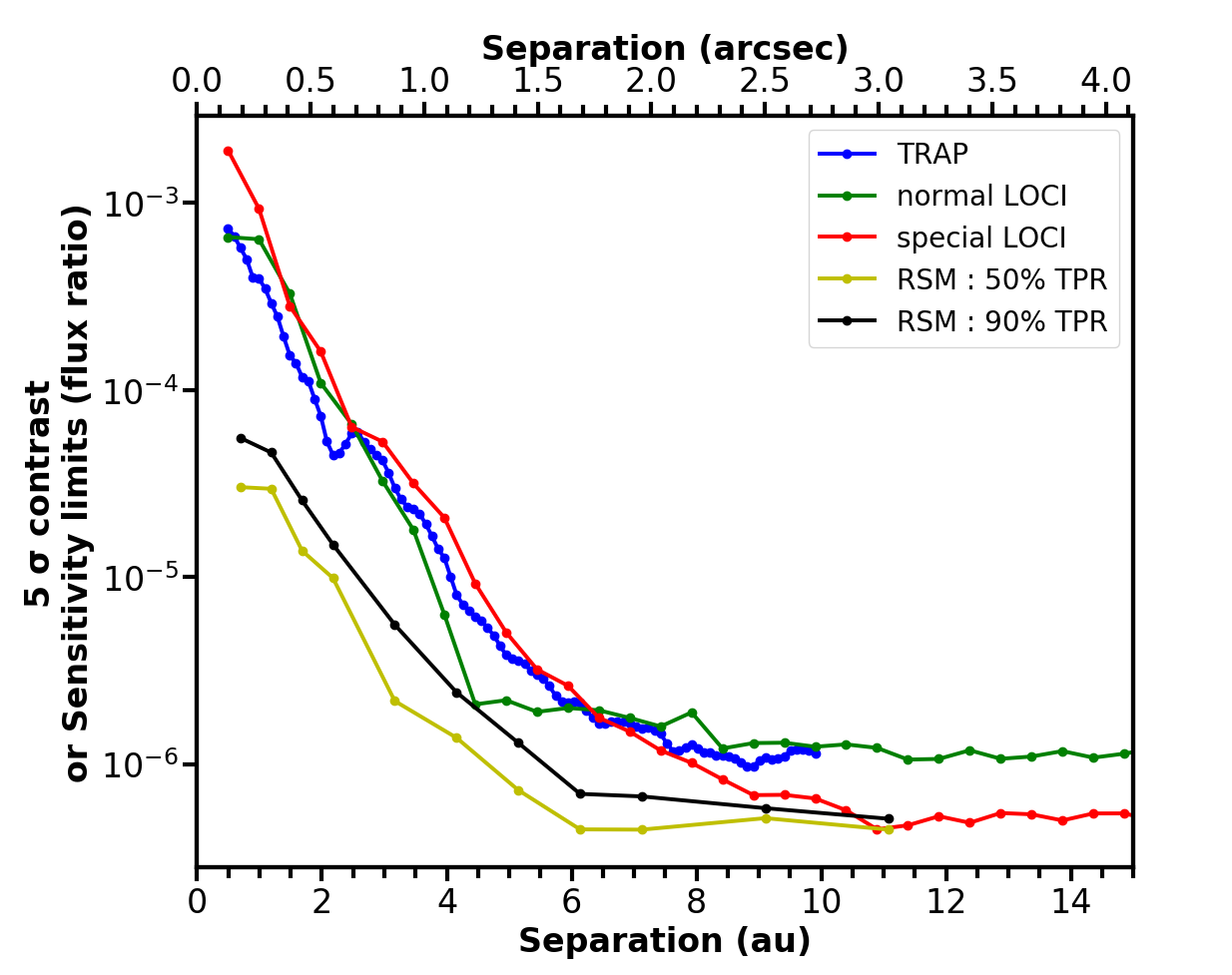}
  \caption{The contrast curves resulting from the different reduction algorithms applied to NaCo L$^{\prime}$ observations in this work. The y-axis shows the 5$\sigma$ contrast limits obtained from TRAP, the normal LOCI and modified LOCI subtraction techniques (shown as blue, green and red dotted line) and the equivalent flux ratio or sensitivity limits obtained using RSM algorithm at 50\% and 90\% TPR as yellow and black dotted lines respectively.}
  \label{comp_con}
\end{figure}

\begin{figure*}
\centering
\begin{subfigure}{.5\textwidth}
  \centering
  \includegraphics[width=0.96\linewidth]{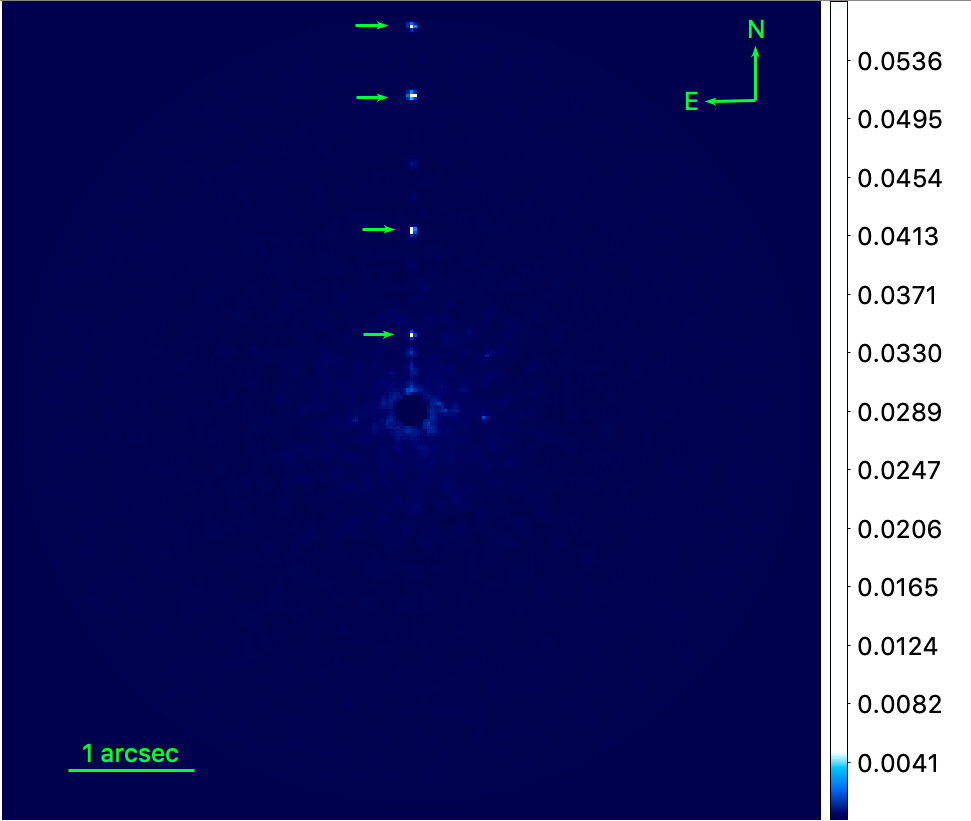}
  \caption{}
  \label{fig2a}
\end{subfigure}%
\begin{subfigure}{.5\textwidth}
  \centering
  \includegraphics[width=1.\linewidth]{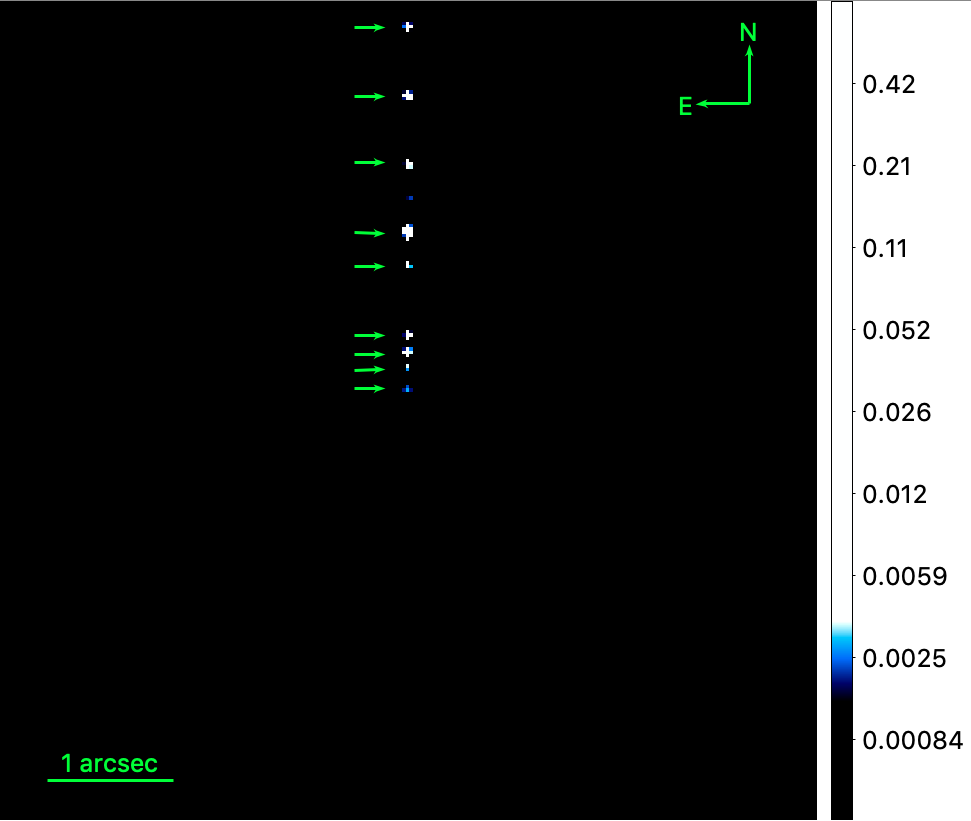}
  \caption{}
  \label{fig2b}
\end{subfigure}
 \caption{The detection maps from NaCo reduction using RSM at (a) 50\% TPR and (b) 90\% TPR. Out of the 11 injected fake companions along the shown axis, 4 have been retrieved for the 50\% case and 9 for the 90\% case. The detected signals correspond to a threshold of 0.0025 which is equal to the brightest speckle in the empty probability map.}
\end{figure*}

The combined contrast curve from both RSM and the modified LOCI procedure is shown in Figure \ref{fig3a}, with the sensitivity limits from RSM shown for both 50\% and 90\% TPR.  

As an alternative check, we also reduced NaCo data by using a combination of the classical LOCI, the modified LOCI and TRAP algorithms. In the classical version of LOCI we shifted the images to a common central pixel location before PSF subtraction, each shift determined through cross-correlation, and performed optimization in an annulus between 40 and 60 pixels of separation from the central star. Self-subtraction was accounted for in the same way as described above for the modified LOCI reduction. As before, in the background limited regime, we performed the modified LOCI implementation where the individual images are not shifted to a common centre before reduction to account for the effects from systematic detector noise. The resulting 5$\sigma$ contrast curve was also derived similar to the modified LOCI reduction. In the regime very close to the star, the LOCI procedures described above exhibit substantial self-subtraction, so we ran the TRAP algorithm \citep{samland2020} on the data, with data from the different nights treated as one contiguous dataset. TRAP being a temporal optimization algorithm, instead of shifting the actual images, the stellar drift in the images was incorporated in the temporal forward model constructed by the algorithm, with the fraction $f$ of principal components set to 0.3. The resulting final output of TRAP showed an improvement in contrast at the very smallest separations (within $\sim$0.6$^{\prime \prime}$), as expected. Hence, we derived contrast curves based on three separation ranges in which three different algorithms provide an optimal result. Consistent with the previous reduction, no companions were detected in the reduced image obtained thus, as well. However, the contrast curves derived in this way were up to an order of magnitude less sensitive than the contrast curve from  RSM algorithm at both 90\% and 50\% TPR (refer Figure \ref{comp_con}). Thus for the reminder of this analysis, we continue with the contrast curves derived from RSM and modified LOCI algorithms (Figure \ref{fig3a}). Conversion into absolute magnitudes was based on WISE \citep{wright2010} photometry, where the W1 filter was used as a proxy for the $L^{\prime}$-band ($m_{W1}=2.9$).

\subsection{NEAR reduction}
\label{s:datanear}

The science data from NEAR observations of $\epsilon$ Ind A, after binning, were available as chopping subtracted and nodding-corrected frames. Since chopping takes care of background as well as any bias/dark current in the data and nodding cleans up any residuals from change in the optical path through the telescope and instrument due to chopping, no additional measures were taken to correct for such effects. Following \citet{petit2020}, no flat fielding corrections were applied to the data, as they do not improve the performance for NEAR data. The stellar PSF subtraction was thus performed on the 223 nod-difference frames stacked into three cubes, one for each observation night.

At the time of observation, the expected planet-star separation for $\epsilon$ Ind system was merely $1.02\arcsec$ ($\sim3.6\lambda/D$). Since our observations operate at 10—12.5 $\mu$m, the diffraction-limited PSF core (0.28$\arcsec$) is much wider than for high-contrast imaging at shorter wavelengths. At the small separations we are interested in, ADI-based reductions schemes are therefore impractical, since the self-subtraction imposed by such techniques become excessively large for modest amounts of field rotation. We thus adopted a recently introduced reduction technique called circularized PSF subtraction \citep{petit2020}, which is independent of field rotation and hence efficient for detecting companion signals at smaller angular separations. This technique involves creating a circularly symmetric stellar PSF by rotating the data frames through 360\si{\degree} in steps of 1\si{\degree} and taking the average of all such rotated versions of the frame. The stellar PSF thus created for each data frame is then subtracted from the original frame and the resulting frames are de-rotated to align the North upwards and East to the left. The final reduced image is then obtained by combining all the de-rotated reduced frames into a master image by means of a weighted sigma-clipped median function, choosing a threshold of 3$\sigma$, with the weights based on the standard deviation of each image. The circularized PSF subtraction technique is effectively equivalent to the classical approach of circular profile subtraction \citep{lafreniere2007}, which has been commonly used in the past to reduce imaging data. In this reduction technique, the image is divided of into a sequence of narrow circles centered on the star. The mean (or median) of each circular area is then calculated and subtracted from all pixels within that area. We have tested both approaches and found that they are indeed essentially equivalent. However, in addition to circular profile subtraction being computationally faster, in this circumstance, it also has a marginally better performance at the specific expected separation on $\epsilon$ Ind Ab, so we use it as the primary option for this study.

\begin{figure}
  \centering
  \includegraphics[width=1.05\linewidth]{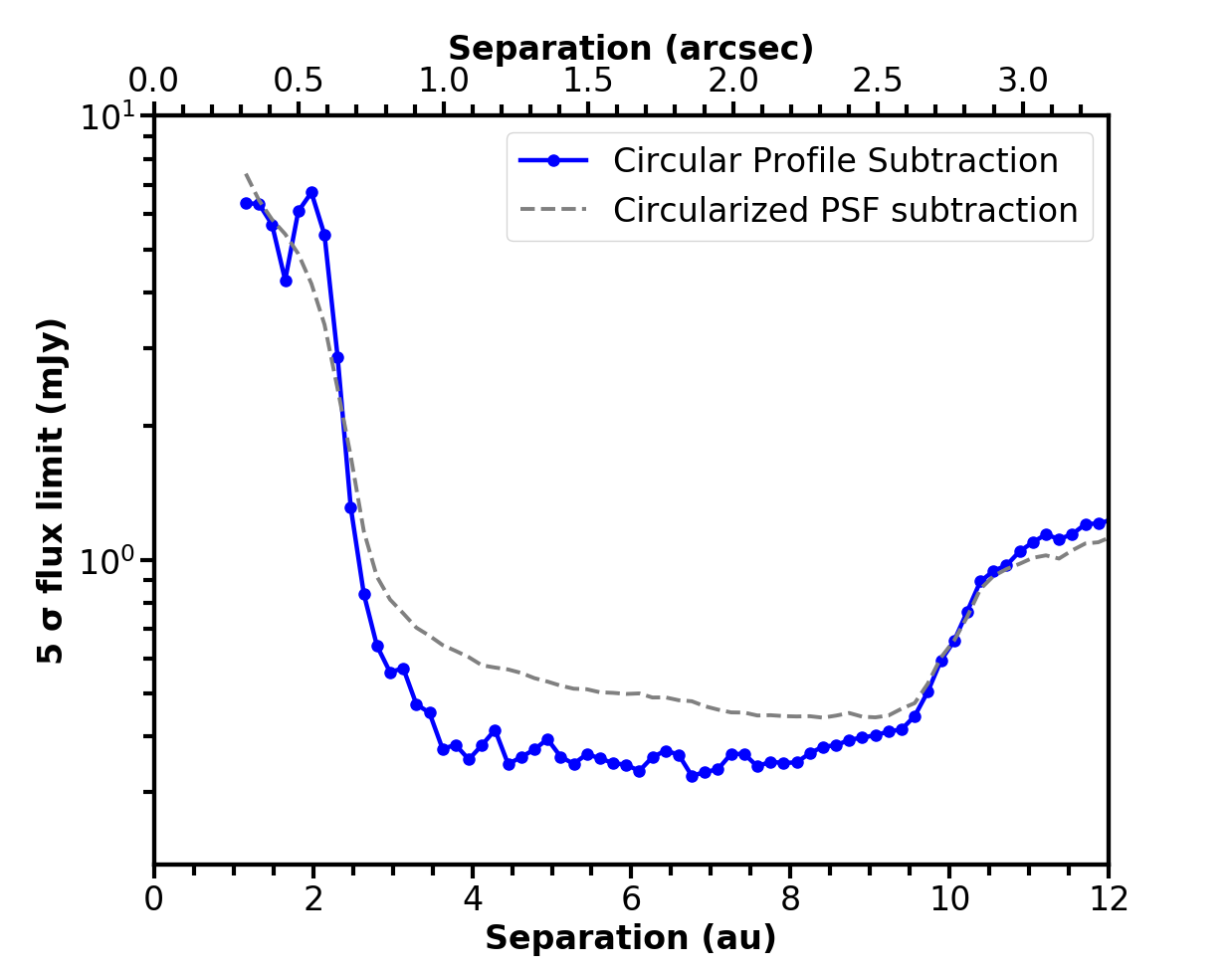}
  \caption{The 5$\sigma$ upper flux limits from NEAR data reduction, obtained using the classical circular profile subtraction technique (blue solid line) and circularized PSF subtraction technique (grey dashed line). The flux limits from both techniques become worse beyond 10 au due to the stellar residuals from chopping/nodding features in the image.}
  \label{fig3b}
\end{figure}

The result of the NEAR reduction is shown in Figure \ref{fig1b}. The expected position of the companion with respect to the central star, as predicted from the best-fit ephemerides for the observation epoch, is also shown in the figure. As with NaCo data, no companion was directly visible in the final reduced image from the NEAR observation. The 5$\sigma$ upper flux limits at different separations from the central star were calculated by fitting Gaussians at each such annulus with an FWHM of 1$\lambda/D$ and a peak equal to 5 times the standard deviation at that location. The photometric reference value used for this flux calculation was 5.68($\pm5\%$) Jy, which is the flux in 11.6~$\mu$m IRAS:12 band for $\epsilon$ Ind A as given in the VizieR photometry tool. The resulting sensitivity curve from the reduced image for NEAR observation is shown in Figure \ref{fig3b}, both from circularized PSF subtraction and the classical circular profile subtraction, with the latter being used for further analysis in this work. As can be predicted from the reduced image from Figure \ref{fig1b}, negative residual images of the star from the nodding/chopping procedure worsen the contrast beyond $\sim$10 au. However, we do not expect this to affect our results, since the range of projected planet-star separation covered at NaCo and NEAR observation epochs from best-fit ephemerides is between 3.71 au (1.02$\arcsec$) and 4.98 au (1.37$\arcsec$).

As an alternative procedure, we also reduced the NEAR data using the TRAP algorithm, keeping in mind its usefulness at smaller separations relative to conventional ADI techniques. TRAP was run on a stacked dataset of all the frames from the three observation nights. The principal component fraction \textit{f} for reduction was set to 0.3 just as in the case for NaCo reduction.  However, the contrast obtained from TRAP reduction was less deep than that obtained via the classical circular profile subtraction technique in the range of separation we are interested in. Hence, for the work in this paper, we use the upper flux limits and contrast obtained via the latter technique for analysis. We account for attenuation in the obtained upper flux limits due to AGPM off-axis transmission in NEAR using the transmission curve provided in \cite{maire2020}. The conversion of the resulting contrast to absolute magnitudes was calculated using the apparent magnitude of $\epsilon$ Ind A in the mid-infrared band WISE W3 (12.082~$\mu$m), $m_{W3}=2.146$, as a proxy to NEAR 10-12.5~$\mu$m band and the distance modulus for $\epsilon$ Ind A, $\mu=-2.1951$.

\section{Results and discussion}
\label{s:results}
\begin{figure*}
\centering
\begin{subfigure}{.5\textwidth}
  \centering
  \includegraphics[width=1.05\linewidth]{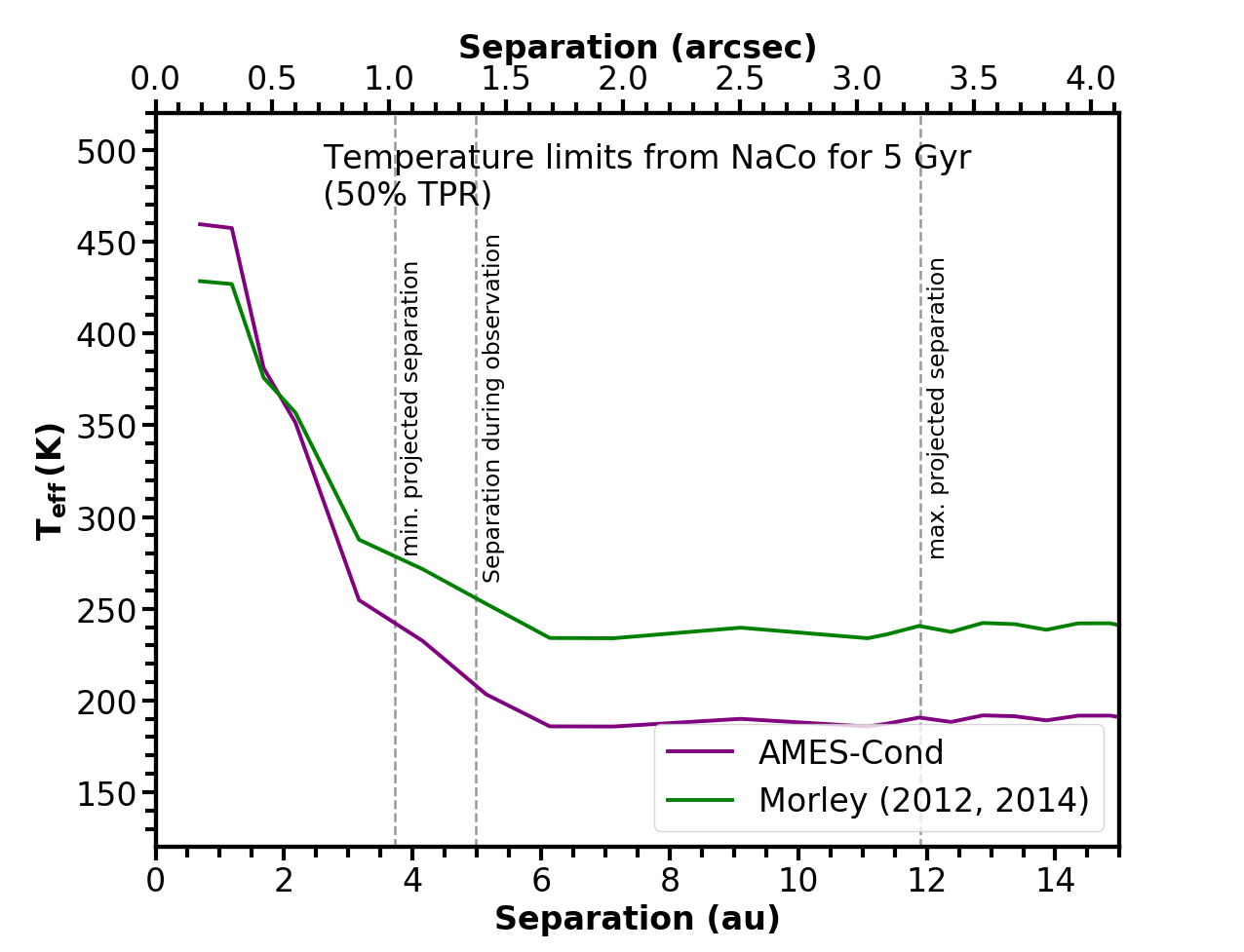}
  \caption{}
  \label{fig4a}
\end{subfigure}%
\begin{subfigure}{.5\textwidth}
  \centering
  \includegraphics[width=1.05\linewidth]{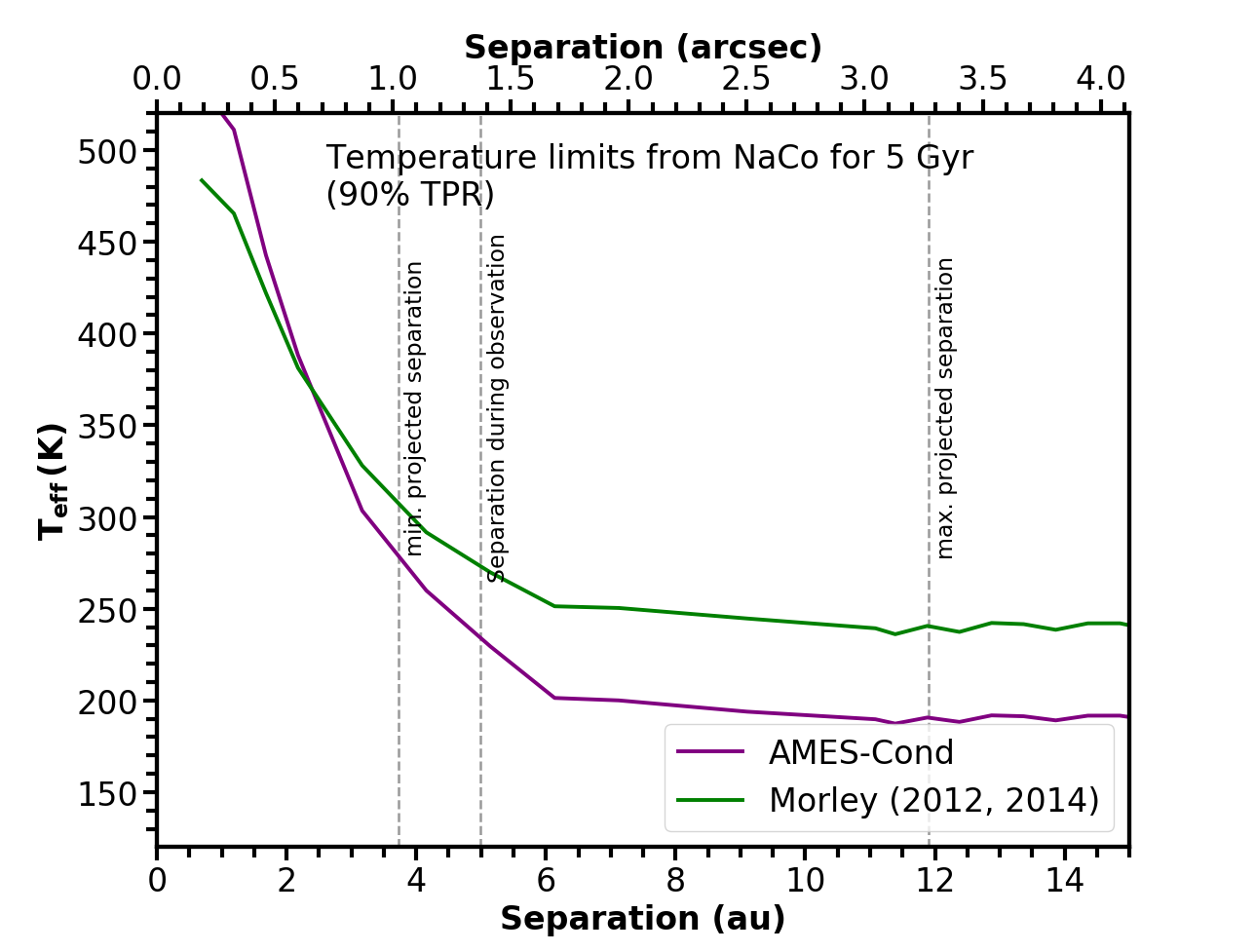}
  \caption{}
  \label{fig4b}
\end{subfigure}
 \caption{The predicted 5$\sigma$ temperature limits for $\epsilon$ Ind Ab from NaCo from contrast curves derived at (a) 50\% TPR and (b) 90\% TPR. The limits derived from the atmospheric models in AMES-Cond grids is shown as solid purple line and the limits from the \cite{morley2012, morley2014} atmospheric models is shown as solid green line. The age assumed for the above limits is 5 Gyr. The dashed black lines represent the minimum, maximum projected separation of the $\epsilon$ Ind Ab as well as the specific projected separation at the epoch of observation.}
\end{figure*}

We derive mass detection limits for planetary companions as a function of distance from the central star from the obtained NaCo and NEAR magnitude limits using theoretical models for predicting the mass-luminosity relationships at different ages based on the brightness in each respective photometric band. For the purpose of our analysis, we used two different sets of models: The AMES-Cond \citep{allard2001, baraffe2003} atmospheric and evolutionary model grids developed for T-dwarf and giant planet atmospheres, and the \citet{morley2012, morley2014} atmospheric models developed for T/Y dwarfs and Y dwarfs. AMES-Cond grids couple atmospheric models with evolutionary models to predict the evolution of effective temperature and luminosity as a function of age for different masses. The atmospheric model described by AMES-Cond grids assumes the dust grains to have settled down gravitationally below the photosphere, hence neglecting the dust opacity in the radiative transfer equation. These models are suitable for \teff~$<1400$K and are known to agree well with NIR photometry beyond 1 $\mu$m. We used the AMES-Cond grids to convert NaCo L$'$ absolute magnitudes to \teff~and mass (\Mp) limits assuming ages of 1, 2, 3, 4 and 5 Gyrs. The temperature limits predicted by the AMES-Cond grids are very similar for all assumed ages from 1 to 5 Gyrs; as an example, the predicted detectable \teff~for 5 Gyr as a function of increasing distance from the central star is shown in Figures \ref{fig4a} and \ref{fig4b} corresponding to contrast curves at 50\% and 90\% TPR respectively. In this and subsequent figures, we also show the minimum (1.02$\arcsec$) and maximum (3.27$\arcsec$) projected separation for the orbit of $\epsilon$ Ind Ab as determined by \citet{feng2019}, as well as the specific predicted separation at the epoch of the NaCo observations (1.37$\arcsec\pm0.16\arcsec$, at a position of <ra>=1.311$\arcsec\pm0.179\arcsec$, <dec>$=-0.061\arcsec\pm0.392\arcsec$, relative to the star). 

The corresponding mass detection limits at 50\% and 90\% TPR are shown in Figures \ref{fig6a} and \ref{fig6b} respectively, as a function of distance from the central star, for age assumptions of 1-5 Gyr. The companion mass predicted from \cite{feng2019}, $3.25^{+0.39}_{-0.65}$ \mj~is also shown in the figures. Assuming an age of 1 Gyr for the system, at the predicted distance of 1.37$\arcsec$ from the central star, the AMES-Cond grids predict a 5$\sigma$ detection limit of \teff~= 207 K and \Mp~= 1.8 \mj~ (M$_{L'}=20.3$ mag) at 50\% TPR and \teff~= 227 K and \Mp~= 2.1 \mj~ (M$_{L'}=19.7$ mag) at 90\% TPR. At an age of 5 Gyr, the 5$\sigma$ limits from \text{the grid} predictions at 1.37$\arcsec$ are \teff~= 208 K and \Mp~= 4.8 \mj~at 50\% TPR and \teff~= 234 K and \Mp~= 5.9 \mj~at 90\% TPR. For both 50\% and 90\% TPRs, the obtained mass limits from the AMES-Cond grids indicate that at an age of 1 Gyr, the companion should be detectable in NaCo $L^{\prime}$ throughout its orbit. At 5 Gyr, $\epsilon$ Ind Ab does not fall within the mass detection limits in its range of projected orbital separation from the central star, at both 50\% and 90\% TPRs.

In atmospheric models like the one described by the AMES-Cond grids, the planet atmospheres are assumed to be free of dust opacity (and thereby clouds), the flux is predicted to arise in relatively deep layers of the atmosphere. By contrast, when cloud opacity is included in the radiative transfer equation, the depth into the atmosphere is restricted, thereby limiting the observed flux particularly at short NIR wavelengths. The \citet{morley2012, morley2014} atmospheric models include cloud opacity in the radiative-convective equilibrium model of brown dwarf atmospheres, with the \cite{morley2012} grid suitable for T/Y dwarfs with surface temperatures between 400-1200 K and log(\textsl{g}) between 4 to 5.5; and the \cite{morley2014} grid suitable for Y dwarfs with surface temperatures between 200-450 K and log(\textsl{g}) between 3 to 5. In addition to the cloud opacity for Na$_2$S, KCL, ZnS, MnS and Cr modelled in \cite{morley2012}, \cite{morley2014} also includes H$_2$O and NH$_3$ that become significant condensates in the atmospheres of brown dwarfs cooler than 500 K. To consider the effect of clouds on the predicted mass limits, we use the \cite{morley2012} atmospheric model for the background limited regime and \cite{morley2014} atmospheric model for the contrast limited regime in our reduced NaCo data. Further, we constrain both these atmospheric models to solar metallicity and a log(\textsl{g}) = 4.0, which is the typical surface gravity of giant planets in the expected mass range as $\epsilon$ Ind Ab. The model is also described by a sedimentation efficiency parameter, \textit{f}$_{sed}$,  that decides the total amount of condensates assumed in each layer of atmosphere, directly affecting the predicted flux described by the model via Mie scattering. A higher \textit{f}$_{sed}$ factor describes optically thinner clouds in the atmosphere, i.e. lesser vertical extension of the clouds and larger particle sizes. In this work, we assume moderate optical thickness for the clouds and set the \textit{f}$_{sed}$ to a value of 5.0 in the model. Using the resulting constrained atmospheric model and based on the absolute NaCo L$'$ magnitudes obtained, we then predict a \teff~for the companion at increasing orbital separations at both 50\% and 90\% TPRs, as shown in Figures \ref{fig4a} and \ref{fig4b}.

In order to utilize the same evolutionary track as the AMES-Cond atmospheric and evolutionary model grids for the purpose of comparing the final mass limits, we use the predicted \teff~by \citet{morley2012,morley2014} atmospheric models to obtain a corresponding mass limit for the companion from the evolutionary grid underlying the AMES-Cond grids, assuming ages of 1, 2, 3, 4 and 5 Gyrs. The corresponding log(\textsl{g}) values obtained this way do not have any significant offset from the log(\textsl{g}) to which we constrain the Morley models. The mass detection limits thus obtained from contrast curves at 50\% and 90\% TPRs are shown in Figures \ref{fig6a} and \ref{fig6b} respectively. At 1 Gyr, at the predicted planet-star separation of 1.37$\arcsec$, the Morley atmospheric models, coupled with evolutionary track from AMES-Cond grids, predict a 5$\sigma$ detection limit of \teff~= 256 K and a mass \Mp~= 2.8 \mj~ at 50\% TPR and \teff~= 273 K and a mass \Mp~= 3.25 \mj~at 90\% TPR for the planet. Assuming an age of 5 Gyr, at 1.37$\arcsec$, the predicted \teff~and \Mp~are 256 K and 6.9 \mj~ at 50\% TPR and 273 K and 7.8 \mj~at 90\% TPR. Based on the derived detection limits from \citet{morley2012,morley2014}, $\epsilon$ Ind Ab should be detectable in NaCo L$^{\prime}$ beyond 1.1$\arcsec$ (4 AU) at 50\% TPR and beyond 1.37$\arcsec$ (4.98 AU) at 90\% TPR if the assumed age for the system is 1 Gyr. Similar to AMES-Cond grids predictions, the Morley grids also predict that the companion is undetectable in the NaCo L$^{\prime}$ band within its projected separation range for an assumed age of 5 Gyr, for both 50\% and 90\% TPRs.

Table \ref{tab1} summarizes the $5\sigma$ \teff, \Mp~detection limits for the companion obtained from NaCo L$^{\prime}$ observations of $\epsilon$ Ind A for assumed age of 1 and 5 Gyr. The detectable mass limits from these observations are a significant improvement from those of previous imaging campaigns of $\epsilon$ Ind A; In comparison, at an age of 1 Gyr, using the \cite{baraffe2003} models, \cite{geissler2007} arrived at a detection limit of 21 \mj~ at separations of $\geq$1.3$\arcsec$ and $16\pm4$ \mj~at $\geq3\arcsec$ from observations in the NaCo H and Ks bands, while \cite{janson2009} arrived at a constraint of 5--20 \mj~for the companion at separations of 10--20 AU ($\sim$2.7$\arcsec$-5.5$\arcsec$) from observations in the NaCo NB4.05 narrow-band and L$^{\prime}$ band.

\begin{figure}
  \centering
  \includegraphics[width=1.05\linewidth]{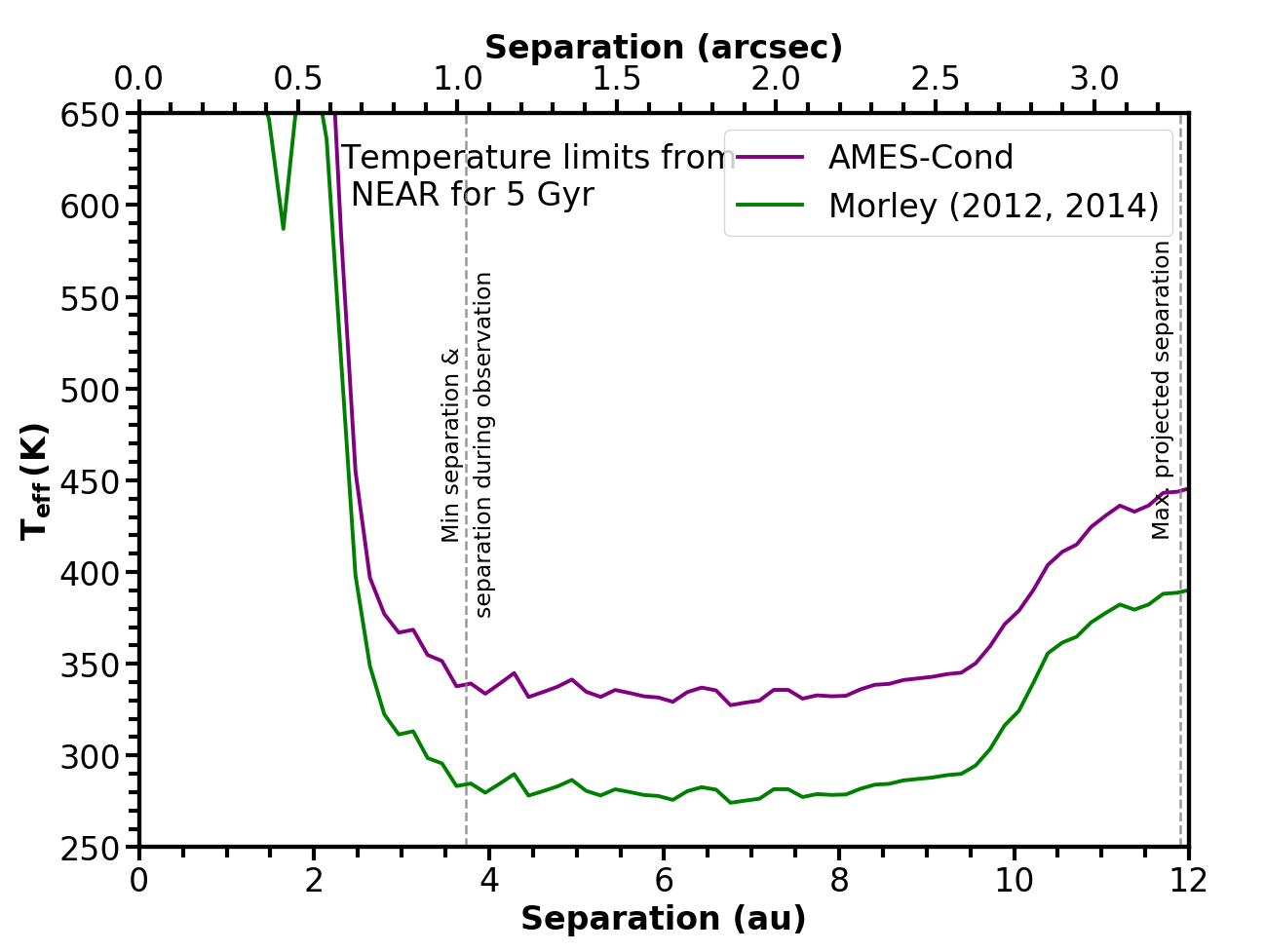}
  \caption{The predicted 5$\sigma$ temperature limits for NEAR magnitudes as a function of increasing distance from the central star. The limits derived from AMES-Cond atmospheric and evolutionary grids is shown as solid purple line and the limits from \cite{morley2012, morley2014} atmospheric models is shown as solid green line. The age assumed for the above limits is 5 Gyr. The dashed black lines have the same meaning as in the previous figures.}
  \label{fig5}
\end{figure}%

\begin{figure*}
\centering
\begin{subfigure}{.5\textwidth}
  \centering
  \includegraphics[width=1.05\linewidth]{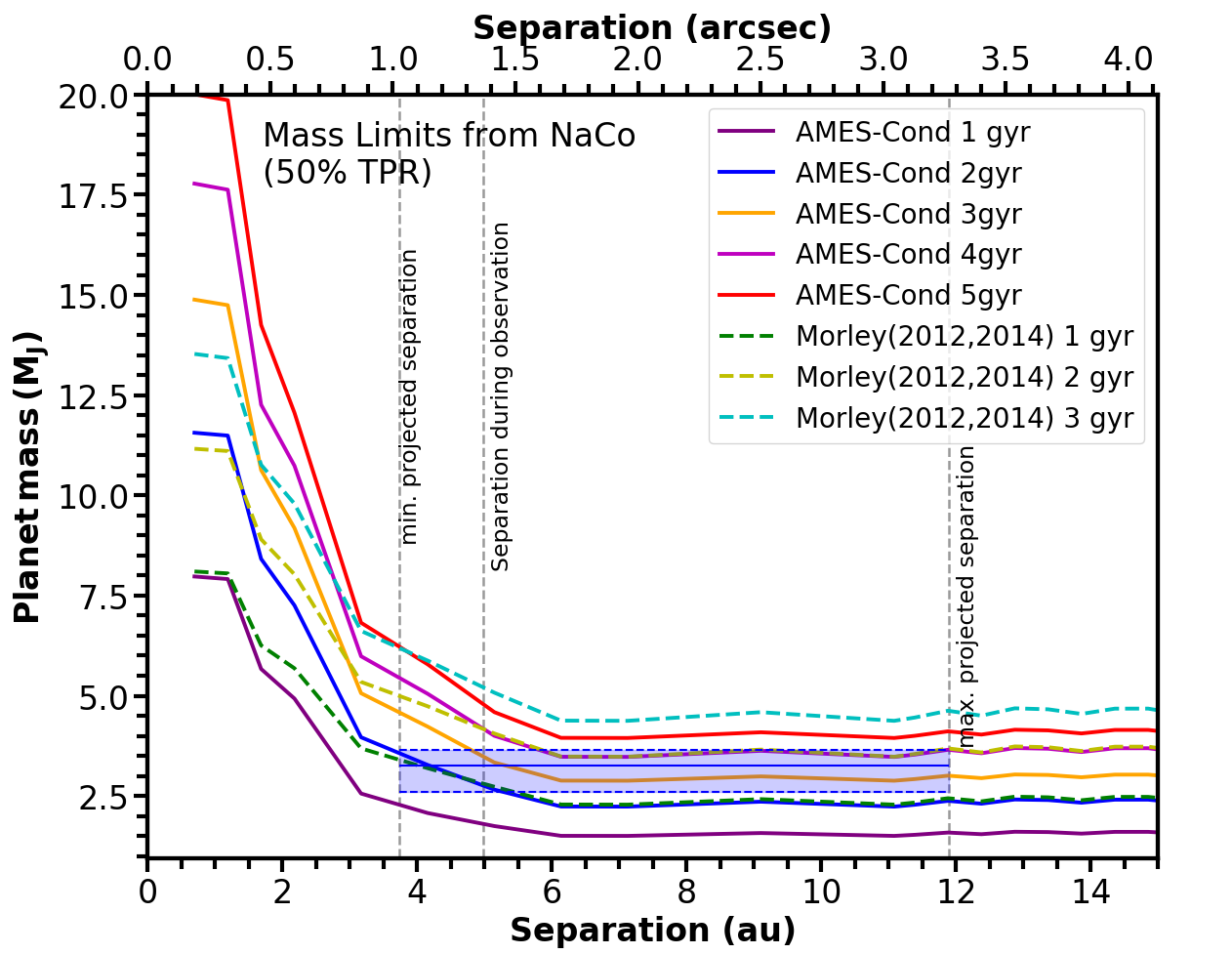}
  \caption{}
  \label{fig6a}
\end{subfigure}%
\begin{subfigure}{.5\textwidth}
  \centering
  \includegraphics[width=1.05\linewidth]{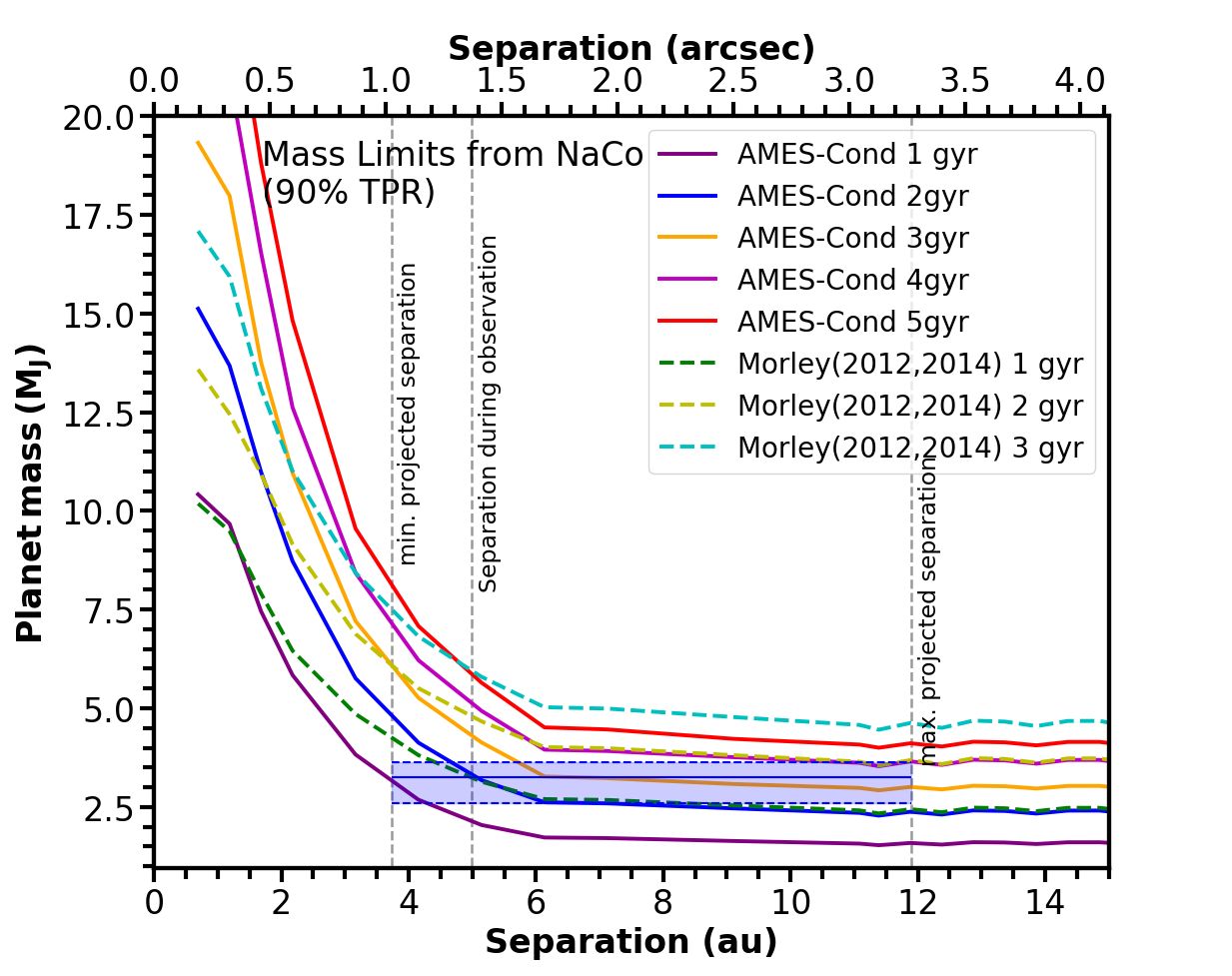}
  \caption{}
  \label{fig6b}
\end{subfigure}
 \caption{The 5$\sigma$ upper mass limits for $\mathbf{\epsilon}$ Ind Ab derived from observations with NaCo $L^{\prime}$, reduced at a TPR of (a) 50\% and (b) 90 \%, for different age assumptions. The solid lines represent limits obtained from AMES-Cond atmospheric and evolutionary grids and the dashed lines represent the limits from \citet[]{morley2012,morley2014} atmospheric models, coupled with the evolutionary track from AMES-Cond grids. The solid blue line represents the companion mass predicted from \cite{feng2019} at an inclination of ${64.25^{\circ}}^{+13.8}_{-6.09}$, with the associated region of uncertainty in mass shown as the filled region between the blue dashed lines. The dashed black lines have the same meaning as in the previous figures.}
\end{figure*}

\begin{table*}[]
\centering
\resizebox{\textwidth}{!}{%
\begin{tabular}{|l|c|c|c|c|}
\hline
Model, Assumed age &
  \begin{tabular}[c]{@{}c@{}}\teff~[K] at 1.5 \arcsec\\ (TPR 50\%, 90\%)\end{tabular} &
  \begin{tabular}[c]{@{}c@{}}\Mp~[\mj] at 1.5\arcsec\\ (TPR 50\%, 90\%)\end{tabular} &
  \begin{tabular}[c]{@{}c@{}}\teff~[K] at 3 \arcsec\\ (TPR 50\%, 90\%)\end{tabular} &
  \begin{tabular}[c]{@{}c@{}}\Mp~[\mj] at 3\arcsec\\ (TPR 50\%, 90\%)\end{tabular} \\ \hline
AMES-Cond 2000, 1 Gyr            & 199, 216 & 1.7, 1.9 & 189, 193 & 1.5, 1.6 \\ \hline
Morley et al.(2012, 2014), 1 Gyr & 247, 264 & 2.6, 3   & 234, 240 & 2.3, 2.4 \\ \hline
AMES-Cond 2000, 5 Gyr            & 198, 220 & 4.4, 5.3 & 186, 190 & 4, 4.1   \\ \hline
Morley et al.(2012, 2014), 5 Gyr & 247, 264 & 6.4, 7.3 & 234, 240 & 5.9, 6.1 \\ \hline
\end{tabular}%
}
\caption{$5\sigma$ \teff, \Mp~detection limits at 1 and 5 Gyr obtained for $\epsilon$ Ind Ab from NaCo L$^{\prime}$ observations, at a separation of 1.5$\arcsec$ and 3$\arcsec$ from the central star, using RSM contrast curves derived at a TPR of 50\% and 90\%.}
\label{tab1}
\end{table*}

%%%%%%%%%%%%%%%%%%%%%%%%%%%
%NEAR Teff and mass limits: AMES-COnd + Morley
%%%%%%%%%%%%%%%%%%%%%%%%%%%

We also apply the AMES-Cond atmospheric and evolutionary model grids and \citet{morley2012, morley2014} atmospheric models to NEAR data to derive detectable \teff~ and mass limits for $\epsilon$ Ind Ab. Since the AMES-Cond atmospheric and isochronal grids currently available for VISIR photometric bands do not include the NEAR band, we calculate the brightness in this bandpass using the theoretical spectra for AMES-Cond available at the Theoretical Spectra Web Server developed by Spanish Virtual Observatory (SVO) \citep{bayo2008}. We constrain the spectra to solar metallicities and a surface gravity of log(\textsl{g}) = 4.0. The NEAR band fluxes for ages 1, 2, 3, 4 and 5 Gyrs were obtained from the spectra by integrating it within the 10-12.5 $\mu$m bandpass and scaling the results by a factor of $\nicefrac{R^2}{D^2}$. Here, D is the distance to $\epsilon$ Ind, and R is the planet radius corresponding to the spectral temperature obtained by interpolating the existing AMES-Cond grid for each respective age. To convert the obtained flux into magnitudes, we use a low-resolution mid-infrared spectrum of Vega from CASSIS \citep{lebo2011} and integrate it over the NEAR band. We use this magnitude calibration and the distance modulus for $\epsilon$ Ind to calculate the absolute magnitudes in the NEAR band from the corresponding flux. We then interpolate the \teff~ and NEAR magnitudes thus derived from the AMES-Cond spectra into the grid points of the regular AMES-Cond grid. We use these new grid values to predict the detectable \teff~ limits for $\epsilon$ Ind Ab in the NEAR band, with Figure \ref{fig5} showing the \teff~ limits with increasing orbital separation for an assumed age of 5 Gyr.

\begin{figure}
  \centering
  \includegraphics[width=1.05\linewidth]{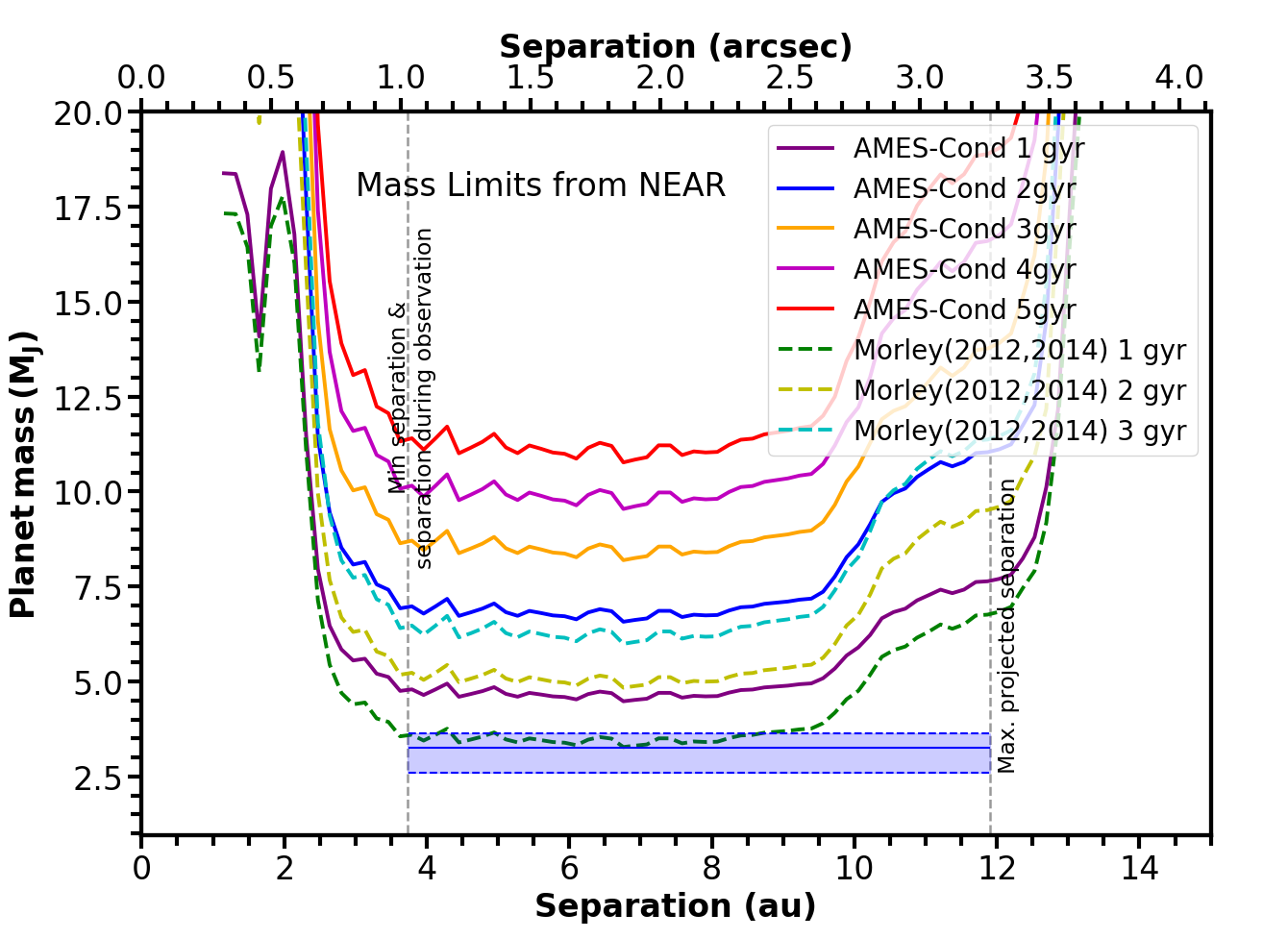}
  \caption{The 5$\sigma$ upper mass limits for $\mathbf{\epsilon}$ Ind Ab derived from NEAR band observations for different age assumptions. The symbols and colors have the same meaning as in the previous figures.}
  \label{fig7}
\end{figure}%

The corresponding upper mass limits obtained for the companion in the NEAR band is shown in Figure \ref{fig7} as a function of orbital separation, for assumed ages of 1 to 5 Gyr. For an age of 1 Gyr, the AMES-Cond atmospheric and evolutionary model grids predict a 5$\sigma$ detection limit of \teff~= 325 K and \Mp~= 4.8 \mj~at the predicted planet-star separation of 1.02$\arcsec$ in the NEAR epoch and for 5 Gyr, the 5$\sigma$ limits obtained from the model at the same separation are \teff~= 338 K and \Mp~= 11.4 \mj. According to the predictions from AMES-Cond grids, the companion does not fall within the detectable range of \teff~and mass in NEAR band observations.

Similar to the analysis of NaCo data, we also apply \citet{morley2012, morley2014} atmospheric models to NEAR data to see how the limits change when clouds are included in the atmosphere. The current Morley grids do not include NEAR band magnitudes, and since the corresponding spectral files were unavailable, we chose the grid for WISE W3 band (12 $\mu$m) as a proxy for the NEAR band. Assuming ages of 1, 2, 3, 4 and 5 Gyr for the system, we predict the detectable \teff~ limits from the obtained NEAR band magnitudes using the corresponding Morley grid. Figure \ref{fig5} shows the 5$\sigma$ \teff~ limits for the planet with increasing orbital distance thus obtained from the Morley atmospheric model for an assumed age of 5 Gyr. As with NaCo data, we then use the evolutionary track for NEAR band AMES-Cond atmospheric and evolutionary model grids for the respective age to predict the final mass limits from the obtained \teff~ limits. The derived upper mass limits for the planetary companion in the NEAR band using the Morley atmospheric model, coupled with evolutionary track from AMES-Cond grids, for different age assumptions is shown in Figure \ref{fig7}. At 1 Gyr, the Morley grids predict a 5$\sigma$ detection limit of \teff~= 284 K and \Mp~= 3.6 \mj~at a separation of 1.02$\arcsec$ from the star and at 5 Gyr, the predicted limits at the same separation are \teff~= 284 K and \Mp~= 8.4 \mj~. In the same way as inferred from AMES-Cond limits, the Morley grids also predict that the expected mass of the companion, 3.25 \mj, is below the detectable mass range for NEAR observations. Table \ref{tab2} summarizes the $5\sigma$ \teff, \Mp~detection limits for the companion obtained from NEAR 10--12.5 $\mu$m observations of $\epsilon$ Ind A for assumed age of 1 and 5 Gyr.

\begin{table*}[]
\centering
\resizebox{\textwidth}{!}{%
\begin{tabular}{|l|c|c|c|c|}
\hline
Model, Assumed age & \teff~[K]  at 1.5" & \Mp~[\mj]   at 1.5" & \teff~[K]  at 3" & \Mp~[\mj]  at 3" \\ \hline
AMES-Cond 2000, 1 Gyr            & 322 & 4.7  & 399 & 7.2  \\ \hline
Morley et al.(2012, 2014), 1 Gyr & 281 & 3.5    & 374 & 6.2  \\ \hline
AMES-Cond 2000, 5 Gyr            & 335 & 11.2 & 426 & 17.6 \\ \hline
Morley et al.(2012, 2014), 5 Gyr & 281 & 8.2  & 374 & 13.6   \\ \hline
\end{tabular}%
}
\caption{$5\sigma$ \teff, \Mp~detection limits at 1 and 5 Gyr obtained for $\epsilon$ Ind Ab from NEAR (10--12.5$\mu$m) observations, at a separation of 1.5$\arcsec$ and 3$\arcsec$ from the central star, using contrast curve from circular profile subtraction technique.}
\label{tab2}
\end{table*}

An interesting point to note from the above results is how in NaCo band predictions from the atmospheric models in AMES-Cond grids for \teff~and \Mp~are lower, hence better in this context, than the respective predictions from  \citet{morley2012, morley2014} atmospheric models for the same age, while in the NEAR band Morley grids predict lower \teff~and \Mp~than AMES-Cond grids. This is a consequence of the \cite{morley2012, morley2014} atmospheric models being a bit redder at relevant temperatures, hence predicting higher flux at longer wavelengths than atmospheric models underlying AMES-Cond grids for the same temperature. This underlines the significance of using different models to predict the \teff~and mass limits of the companion to account for such model uncertainties while interpreting the results. Furthermore, it underlines the value of observing in multiple carefully selected wavelength bands as in this study, since this provides redundancy against the existing model uncertainties, and thereby yields much more robust detections or detection limits.

We note that we have also attempted to use the Exo-REM atmospheric model \citep{baudino2015, baudino2017} as a third alternative model interpretation for placing constraints on the \teff~, and hence mass, of $\epsilon$ Ind Ab. The original Exo-REM model is essentially non-cloudy and valid for EGPs with \teff~= 500-2000 K. An upgraded model for Exo-REM, \cite{charnay2018}, includes both absorption and scattering by clouds in the atmosphere and is valid for EGPs and brown dwarfs with \teff~= 300-1800 K. However, as it turned out during our analysis, the detectable \teff~drops below 300 K for much of the separation range of $\epsilon$ Ind Ab. Since the Exo-REM model grids do not extend to such low temperatures, they are not applicable for the purpose of this analysis. Hence, we do not include them, but we do note that in the temperature range over which they overlap with the other tested model grids, both the cloudy and cloud-free Exo-REM models show good consistency with the atmospheric models underlying the AMES-Cond grids.

\section{Conclusions}
\label{s:summary}

In this work, we present observations of $\epsilon$ Ind A using deep AO imaging at the VLT with NaCo in the $L^{\prime}$-band and NEAR in a dedicated 10-12.5~$\mu$m band, in order to attempt detection of the planetary companion $\epsilon$ Ind Ab. Assisted by the well-constrained ephemerides and mass of the planet, we can derive stringent constraints on its brightness from the imaging. We arrive at unprecedented sensitivities close to the bright star, with detectable planet temperatures as low as 200–300 K depending on the choice of atmospheric model between AMES-Cond and the \citep{morley2014} models. From the corresponding planet mass detection limits obtained from AMES-Cond atmospheric and evolutionary models, the 3.25 \mj~planet $\epsilon$ Ind Ab would be marginally detectable in the NaCo L$^{\prime}$ images if the (uncertain) age of the $\epsilon$ Ind system is as low as 3 Gyr at a TPR of 50\% and 2 Gyr at a TPR of 90\%. On the other hand, the planet mass detection limits as obtained from \citet{morley2014} models at both 50\% and 90\% TPR suggest that the planet will be detectable if the age of the system was 1 Gyr. As can be inferred from the non-detection of the planet in the images at its predicted location of 1.37$\arcsec$, at 50\% TPR, the AMES-Cond model grids thus hint at an age of 4 Gyr or above and the \cite{morley2014} model grids hint at an age of 2 Gyr or above for the planet. At 90\% TPR, the detectable planet mass and \teff~limits are slightly higher, with AMES-Cond model grids indicating an age of 3 Gyr or above but \cite{morley2014} model grids predicting an age of 2 Gyr or above consistent with the former case. Assuming that these two atmospheric models cover a somewhat accurate depiction of the companion's atmosphere, both these cases definitely put a constraint of 2 Gyr for the lower age limit of the planetary candidate. The non-detection of the planet in these observations and the corresponding age estimation in this work is compatible with the indication of an older age for the system in the literature. The NaCo detection limits are more constraining than the NEAR limits according to both model sets, partly due to the high thermal background in NEAR and partly due to the less favourable separation of the planet at the NEAR epoch. It is also to be noted that while the NaCo data is contrast-limited at the expected position of the companion from the best-fit ephemerides, the NEAR data is essentially background limited as can be seen from the absence of any visible speckle structure around the expected planet position in Figure \ref{fig1b}. A longer integration with NEAR may, hence, bring the contrast in this band further down, but for NaCo L$^{\prime}$ it would not result in any significant improvement. However, while the difference in mass detection limits between NaCo and NEAR is very distinct in the AMES-Cond interpretation, it is much more subtle in the \cite{morley2014} interpretation, which predicts a worse sensitivity relative to AMES-Cond for NaCo, but a better relative sensitivity for NEAR. This underlines the important fact that the sensitivity in mass depends not only on the intrinsic instrumental sensitivity in different bands, but also on the model-predicted flux distribution among those bands. In this regard, having measurements in more than one high-sensitivity band as in our study adds considerable robustness against atmospheric uncertainties in the interpretation of the physical detection limits.

For securing an imaging detection of the planet at any realistic age, only a rather modest improvement in sensitivity (e.g., $\sim$1 mag or less in $L^{\prime }$-band) is required, particularly since the projected separation of the planet is expected to be more favourable for detection over the next few decades. The upcoming ERIS \citep{kenworthy2018} instrument for the VLT is foreseen to be able to deliver the required sensitivity in the $\sim$4~$\mu$m range. The Mid-Infrared Instrument (MIRI) \citep{bouchet2015} onboard the James Webb Space Telescope (JWST), which is currently set to launch in late 2021, offers broadband imaging with coronagraphy in a broad wavelength range of interest (10.6, 11.4, 15.5, 23 $\mu$m filters) at unmatched sensitivity \citep[$\sim1.5\mu$Jy limiting sensitivity at 10$\sigma$ for a 10,000s exposure at 11.3 $\mu$m, with a PSF FWHM of 0.36$\arcsec$;][]{bouchet2015}, aiding a possible detection of $\epsilon$ Ind Ab in MIR. The Mid-Infrared ELT Imager and Spectrograph (METIS) \citep[]{brandl2014,quanz2015,brandl2018}, an instrument on the 39m European Extremely Large Telescope (E-ELT) that is set to see its first light in the late 2020s, will offer mid-infrared imaging with AO and coronagraphy with a very high sensitivity \citep[$\sim$0.1mJy at 10$\sigma$ for a 1 hour exposure at $\sim$11 $\mu$m;][]{brandl2014} and contrast, enabling high-resolution spectroscopy for detailed atmospheric characterization. In addition, further radial velocity monitoring in combination with new astrometric data from future Gaia releases can place tighter constraints on the orbital parameters for $\epsilon$ Ind Ab, further aiding high-precision characterization of a wide range of atmospheric and physical properties for this cold and very nearby giant planet.

\begin{acknowledgements}
M.J. gratefully acknowledges funding from the Knut and Alice Wallenberg Foundation. We made use of CDS and NASA/ADS services for the purpose of this study. Part of this work has been carried out within the framework of the NCCR PlanetS supported by the Swiss National Science Foundation. SPQ and AB acknowledge the financial support from the SNSF. Part of this research received funding from the Fonds de la Recherche Scientifique - FNRS under Grant No F.4504.18, and from the European Research Council (ERC) under the European Union's Horizon 2020 research and innovation programme (grant agreement No 819155). HRAJ acknowledges support from the UK’s Science and Technology Facilities Council ST/R000905/1.
\end{acknowledgements}

\end{document}